## Murburn concept: A facile explanation for oxygen-centered cellular respiration

*Kelath Murali Manoj\**

\*Satyamjayatu: The Science & Ethics Foundation
Kulappully, Shoranur-2 (PO), Palakkad District, Kerala, India-679122.
Email: satyamjayatu@gmail.com; satyamjayatu@yahoo.com

***Abstract:*** Via a concurrently communicated manuscript [Manoj, 2017], I have conclusively debunked the long-standing "Electron Transport Chain – Proton Pump – Chemiosmosis – Rotary ATP Synthesis" (EPCR) explanatory paradigm for mitochondrial oxidative phosphporylation (mOxPhos). Now, the mandate is open to establish a viable explanation for mOxPhos. Several similarities/parallels could be noted between microsomal xenobiotic metabolism (mXM) and mOx-Phos, with respect to the reaction components, microenvironment, experimental observations and mechanistic interpretations. Therefore, it is quite logical to apply the insights from mXM studies in the mOxPhos arena. Herein, a coherent proposal centered on the murburn concept (recently established as a probable rationale for mXM [Manoj et al, 2016d]) is detailed as a viable alternative for explaining the overall phenomenology of mOxPhos. The essential logic of the reaction system is unordered, with an overall one-electron (radical!) paradigm. The bi-membrane system of mitochondrion could be deemed functionally analogous to a natural nuclear reactor. But herein, one-electron species are spontaneously generated, probabilistically regulated/controlled and their potential reactivity harnessed for ATP coupling process in a low water – low proton microcosm. The newly proposed concept aligns with known experimental facts, Ockham's razor and evolutionary perspectives. It also provides answers to key questions regarding the initiation, sustenance and termination of the molecular processes of energy transduction in life. Several evidences/arguments are presented in support of murburn concept and experimental approaches are suggested to further ratify or falsify the newly proposed hypothesis. Furthermore, forthright projections discuss the impact of murburn concept on cellular physiology.

Cellular respiration affords aerobic eukaryotes the major chunk of "energy" required for doing the "works" of life. These works of life include macroscopic level activities like response to different stimuli or maintenance of organizational integrity; and molecular level processes such as catalysis or replication. It is evident that these activities involve various interactive mechanical, chemical, thermal, electrical and radiant energies. As a pre-requisite to the fundamental energy generating process, a systematic trimming and gradual stripping of electrons (or hydrogen atoms) of/from the multi-carbon reduced molecule (as exemplified by glucose, fatty/amino acids, etc.) occurs, to give a 2-carbon acetyl moiety. In turn, this acetyl group is taken up by Kreb's cycle, churning out two fully oxidized molecules of $CO_2$. In the overall process, some reduced molecules (like NADH/$FADH_2$) are formed, which are oxidized at the inner mitochondrial membrane, to yield ATP (a ubiquitous energy currency of cellular systems) within the organelle's matrix. All aerobic life forms employ molecular oxygen as the final electron acceptor for the overall chemical process. This "oxidation of reducing equivalents coupled with the synthesis of ATP in mitochondria" serves as the fundamental chemical logic of energy transduction, and is called *oxidative phosphorylation* (OxPhos).

For the past two and a half decades, I could never appreciate the erstwhile explanations for mOx-Phos which relied on highly deterministic and overtly cumbersome "**E**lectron Transport Chain", an energetically improbable relay of "**P**roton Pumps" and "**C**hemiosmosis" based energy transduction by a "**R**otary Synthesis of ATP" by a highly sophisticated and fully reversible trans-membrane enzyme. (I summate this as the EPCR hypothesis.) Via a recent manuscript [Manoj, 2017], I have established the mandate to think beyond the erstwhile explanations. (I believe that it is imperative that the reader peruses the manuscript above before proceeding any further!) Towards that purpose, I propose murburn concept (the ideas originally proposed to explain liver mirosomal xenobiotic metabolism (mXM) [Manoj et al, 2016a-d; Venkatachalam et al, 2016]) as a rationale for the overall phenomenology of mOxPhos. The fundamental treatise of this manuscript is that in cellular respiration, formation of diffusible and reactive oxygen species (DROS) is a mandatory requisite, not an undesired side-reaction (as hitherto perceived!). Given the importance of the subject and the nature/stature of the developments, I hope the ideas proposed herein are given due consideration and deliberation, before any of them are rejected or ratified.

**Results & Discussion**

**1. Similarities between mXM and mOx-Phos:** While working with redox proteins, over the last decade, I have advocated the concept that diffusible reactive species (though, at times chaotic!) can mediate effective and specific catalysis in biological milieu [Manoj et al, 2016a-d; Venkatachalam et al, 2016; Parashar et al, 2014a-b; Gade et al, 2012; Gideon et al, 2012; Parashar & Manoj, 2012; Manoj et al, 2010a-b; Manoj & Hager, 2008; Manoj, 2006]. This simple idea was essentially called "murburn concept" [Manoj et al, 2016a-d; Venkatachalam et al, 2016]. It explained the peroxidative chlorination involved in the seemingly simpler reaction milieu of the fungus *Caldariomyces fumago* (leading to the *in situ* synthesis of the antibiotic caldariomycin) and in the relatively complex enzymatic system found in xenobiotic-metabolizing hepatocytes/ liver microsomes. Murburn concept signifies a "mured burning" or "mild unrestricted burning", connoting a controlled radical (one-electron) reaction scheme involving oxygen and reductant(s). The liver cells' endoplasmic reticulum (microsome) membranes have high concentrations of two hemoproteins, cytochrome P450 (CYP) & cytochrome $b_5$ and low levels of an essential diflavoenzyme, cytochrome P450 reductase (CPR). It is evident from a comparative perspective that the microsomal system offers several parallels and similarities to the mitochondrial system (Table 1). At the outset, it can be seen that both these systems have catalysts with similar cofactors, the reaction system employs reduced nicotinamide nucleotides and oxygen at phospholipid interface, and water / DROS formation is observed in both systems, etc. The distribution densities are also similar- the heme-cofactor containing macromolecules out-numbering the flavin-cofactor containing proteins. As an experimental scientist, I could also note the similarities and parallels between the "classical respiration states of mOxPhos" [Chance & Williams, 1956; Estabrook, 1967; Nicholls & Ferguson, 1992] and the profiles obtained in mXM for the rates of $O_2$ consumption and reduced nucleotide depletion. In mOxPhos research practice, the sequential addition of reaction components into the mitochondrial suspension (followed by tracking of oxygen utilization with oxygen electrode) is structured into 5 "states". Similarly, in the mXM system, negligible rate (of oxygen uptake and NADH depletion) was observed with CYP alone and a slightly higher rate was obtained with CPR alone, which was affected to various levels with a combination of CYP + CPR (depending on the ratios). A much higher rate was obtained when the xenobiotic substrate was added to CYP + CPR mixture and the product formation (rate/yield) depended on the reconstitution methodology and extent of coupling in the system. In both mXM and mOxPhos research, workers had termed such "apparently non-physiological" trickling/wasting of redox equivalents as "electron leaks". I have explained that this is a constitutive redox expense involved in cellular systems, which gets going full throttle when suitable two-electron sinks are readily available. A study of my earlier works [particularly, Manoj et al., 2016d] would show that most of the perceptions in the mXM system (regarding oxygen activation, electron transfers, substrate binding, kinetics, operative thermodynamic logic, etc.) changed owing to a re-interpretation of the phenomenology, as made possible by the murburn concept.

**2. The murburn concept perspective of mOxPhos:** The essential facet of the paradigm I had recently established in mXM is that phospholipid membrane bound flavin system (CPR) generates DROS and heme systems (CYP + Cyt. $b_5$) help stabilize a one-electron reaction paradigm involving DROS (including radicals, peroxide, singlet oxygen, etc.). Therefore, these DROS were deemed to have obligatory and constructive physiological roles (discounting the traditional view that they are primarily disruptive or chaotic!) in electron transfers and specific redox catalysis. The essential components of murburn concept for mOxPhos system are-

*In the physiological redox regime of ~-400 mV to ~+800 mV (a potential range beyond which, water starts breaking up into hydrogen and oxygen gases: $2H^+ + 2e^- \rightarrow H_2$ and $2H_2O \rightarrow O_2 + 4H^+ + 4e^-$), "superoxide radical – peroxide – hydroxyl radical" conversion equilibriums exists in the "oxygen + reductant + Fe (heme)/ flavin" aqueous system. This equilibrium could be modulated by a diverse set of agents like redox proteins and small interfacial redox-active organic molecules (which could modulate DROS dynamics), protons/ions (which could affect the dielectrics and permittivity; thereby affecting the overall reactivities involved).

* In biological systems, there cannot be "high energy electrons or high potential electrons" and it is unlikely that "protons retain energy/potential" just because they move from one macroscopic system to another. (Quite simply put, when an electron or protons changes its address, it retains little information regarding its previous residence!) The events transpire under normal physiological conditions and there is no external heat or electric field input; thereby all electrons (whether they are taken from NADH by Complex I or found at Complex III / Complex IV) in the

system have equal energy. In the newly proposed system, there is no structured/directional flow of electrons from low to high redox potential via tightly connected and/or regulated strings of redox centres. (There could be some transfers that abide by the former understanding, but the "erstwhile flow" is not the primary logic of the new mechanism.)

*Redox-active proteins generate oxidative and reductive 'abilities' in or around their vicinities by virtue of accepting or releasing highly mobile / diffusible one-electron redox-reactive species (or electrons). The outcome of such a process could be either one or two electron stabilized intermediates/products, subject to various redox reactions and partitioning equilibriums. All cofactors involved in electron relay within the "ETC" (from Complexes I through IV- flavins, hemes, Fe-S and Cu-centres, CoQ, Cyt. *c*, etc.) are readily one-electron redox-active species. Positional (presence in matrix / IM / IMS etc.) and structural attributes (hexaligated heme versus pentaligated heme; presence of closely located redox centres, solvent accessibility, etc.) of a particular redox-active species could play significant roles in governing the reaction outcomes.

* Efficient electron relay between proteins and small molecules/ions is spontaneously mediated through highly mobile and redox-active diffusible species within small distances around the vicinity of the phospholipid membrane (murzone). The pertinent small or larger molecules/ions serving as the final "donor-acceptor" combination need not undergo a direct binding (as is currently believed to be essential) and subsequently, long-range electron transfers through the covalent peptide bonds to the redox centres of the proteins are not obligatory. This is because highly diffusible species (which have ~microsecond lifetimes) can possess stability/mobility even up to a few tens of Angstroms around the phospholipid interface.

* In most constitutive metabolic reaction schemes/cycles (of the type A→B→C→D…→X), the major drive for the formation of X is the accumulation of A, B, C, etc. within the reaction milieu/pool. (This is a tentative scenario assuming the lack of feedback inhibitions and allosteric regulations.) Therefore, the protagonists formed in the initial part exerting a forward push forms the major thermodynamic drive for the overall reaction. In my recent work with the one-electron metabolic reaction cycle of cytochrome P450s (CYPs), I had shown that the two-electron reacted products formed at the last reaction stage (formation of X) exerts a thermodynamic pull on the overall one-electron equilibriums (in the initial part of the reaction cycle) to shift towards right.

* The reaction sequence does not seek compartmentalization or multi-molecular ordered/ sequential events. I have already demonstrated that such a one-electron process could be mild,

reproducible, specific/selective and inherently constitutive in nature. The resulting process is simple, functional and affords a viable redox chemical transduction mode in the physiological redox ranges. In the context of mitochondrial reaction system, analogy could be drawn with a nuclear fission reactor (which is in stark contrast to the analogy drawn to the components of an automobile that the debunked EPCR hypothesis sought!). The key difference from the nuclear reactor being that herein, a radical reaction is initiated, moderated, maintained and quenched. (For a detailed discussion, please refer Item 1, Supplementary Information.]

*Summary of the reaction system with holistic perspectives:* A schematic cross-sectional representation of the salient participants involved in mOxPhos is shown in Figure 1a and their surface distribution is shown in Figure 1b. Murzymes (enzymes that mediate unrestricted redox reactions) of various redox potentials, Complexes I, II & III initiate a one-electron reaction from the substrates NADH, succinate and $CoQH_2$ respectively, generating one-electron equivalents in the milieu, which could be (enzymatically or non-enzymatically) relayed to oxygen on the inside, CoQ within the membrane and Cyt. *c* in the IMS of the mitochondrion. The metallic and reducible centers present are not primarily for systematic or sequential relay of the one or two electron equivalents, but to synthesize and stabilize oxygen-centered radical intermediates and to overcome spin barriers. Overall, the reaction may be initiated at several points, if a suitable reducing agent is present within the system as the electron donor. Therefore, this is a confined burning of redox equivalents that would be fast enough to explain the need for high amounts of oxygen and phosphate, which serve as the electron sinks (thermodynamic pull!) to drive the equilibrium towards the right. Therefore, rather than a sequential and ordered distribution of components and mechanism thereof, the actual functionality is distributed and delocalized high potential hotspots. A sustaining paradigm is afforded by further stabilization of superoxide and its reduction to peroxide, leading to the generation of singlet oxygen and hydroxyl radical by Fe/Cu centers of Complex IV. Now, chemical coupling (connecting oxidation of NADH to the phosphorylation of ADP) is mediated by the highly reactive intermediates, around the vicinity of Complex IV & V. Complex V could help coupling in an indirect manner, by sustaining the proton requirement in the system. *In toto*, the earlier conceived rudimentary functionality of the ETC can be at best envisaged as a back-up mechanism to recycle lost electrons within the system and to pull electrons into a two-electron sink in the immediate vicinity of the respective complexes. We do not need to entertain improbable suppositions such as "a molecule of oxygen waits indefinitely for four protons and electrons each, and that Complex IV's reaction centre has an uncanny ability to tether oxygen and its diverse forms until water is formed". Similar misunderstandings were done away in the CYP reaction system, where a highly electron deficient CYP-based Compound I was "mistakenly" supposed to form water. The system is just a series of unordered binary collisions (subject to probabilistic operators) that would occur on/around the inner phospholipid membrane. So, rather than serving as some "willing operator creating an eternal potential difference of protons and electrolytes" across itself, the membrane serves as a controlled burning location involving high-potential reactive radicals. The point to be seen is – in this controlled regime, as long as there are more reactive species, the lesser reactive species don't react (on a statistically relevant scale, until certain thresholds are reached). That is one of the operative and regulatory principles of murburn concept, which explains oxidative stress (at high and low concentrations of oxygen). This setup would not need an ordered placement of components nor would it need any sequential or synchronized process of electron transfer from one protein to the other. Such a system serves Ockham's razor and has a high probably to evolve from a minimal set of components. Very importantly, with the murburn concept, we can now appreciate the possibility of a higher "theoretical" ATP synthetic yield (more than the prevailing limitation of ~50%) with oxidative phosphorylation. The reactions at the initiating Complexes could be represented as-

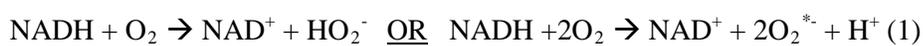

$NADH + O_2 \rightarrow NAD^+ + HO_2^-$  <u>OR</u>  $NADH + 2O_2 \rightarrow NAD^+ + 2O_2^{*-} + H^+$ (1)

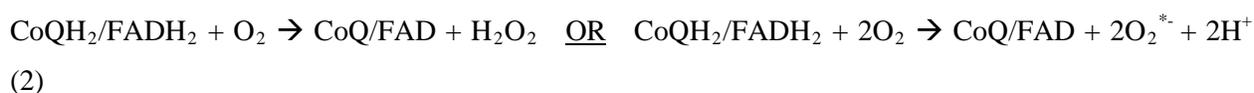

$CoQH_2/FADH_2 + O_2 \rightarrow CoQ/FAD + H_2O_2$  <u>OR</u>  $CoQH_2/FADH_2 + 2O_2 \rightarrow CoQ/FAD + 2O_2^{*-} + 2H^+$ (2)

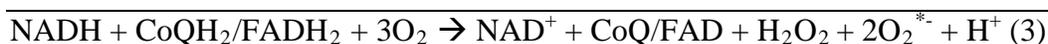

$NADH + CoQH_2/FADH_2 + 3O_2 \rightarrow NAD^+ + CoQ/FAD + H_2O_2 + 2O_2^{*-} + H^+$ (3)

In the milieu (or in Complex IV's vicinity), now-

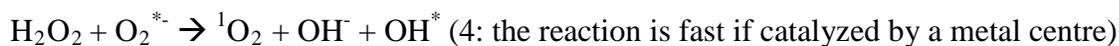

$H_2O_2 + O_2^{*-} \rightarrow {}^1O_2 + OH^- + OH^*$ (4: the reaction is fast if catalyzed by a metal centre)

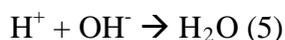

$H^+ + OH^- \rightarrow H_2O$ (5)

Therefore, the overall equation becomes-

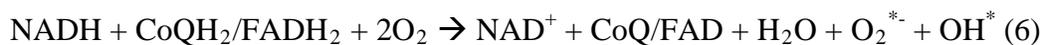

$NADH + CoQH_2/FADH_2 + 2O_2 \rightarrow NAD^+ + CoQ/FAD + H_2O + O_2^{*-} + OH^*$ (6)

Or, for one oxygen molecule, the equation is-

$$NADH + (C_4H_4O_4)^{2-} + O_2 \rightarrow NAD^+ + (C_4H_2O_4)^{2-} + H_2O + OH^- \text{ (7)}$$

Concomitantly, the radicals produced above serve as couplers in the matrix reaction:

$$Ad\text{-}O\text{-}(PO_2)^-\text{-}O\text{-}(PO_3)^{2-} + HPO_4^{2-} + H^+ \rightarrow Ad\text{-}O\text{-}(PO_2)^-\text{-}O\text{-}(PO_2)^-\text{-}O\text{-}(PO_3)^{2-} + H_2O \text{ (8)}$$

As per the murburn concept, the reaction generates hydroxide ions and water. The system does not solicit outward proton pumping machineries at the phospholipid interface. Also, the logic is predominantly a one-electron paradigm that invokes the involvement of oxygen at multiple points (in space and time!) and all reactions in this scheme are bimolecular, brought about in free solution or catalyzed by a protein around the phospholipid interface. The mechanism shows why both NADH (a 2-electron + 1-proton source) and $FADH_2$ (a 2-electron + 2 proton source) are incorporated into the mitochondrion. Besides equation 8, the critical dependence of rates and thermodynamics on pH can be well understood by considering the disproportionation and one-electron redox equilibrium of a key element of murburn concept- superoxide [Kanematsu & Asada, 1994; Sawyer & Valentine, 1981].

Regarding superoxide dismutation kinetics:

$$HO_2^* + HO_2^* \rightarrow O_2 + H_2O_2 \ [k = 1 \times 10^5 \ M^{-1}s^{-1} \text{ (acidic)] (9)}$$

$$HO_2^* + O_2^{*-} (+ H^+) \rightarrow O_2 + H_2O_2 \ [k = 1 \times 10^8 \ M^{-1}s^{-1} \text{ (pH = 4.8)] (10)}$$

$$O_2^{*-} + O_2^{*-} (+ 2H^+) \rightarrow O_2 + H_2O_2 \ [k \leq 3 \times 10^{-1} \ M^{-1}s^{-1} \text{ (alkaline)] (11)}$$

For the reduction of oxygen:

$$O_2 + H^+ + e^- \rightarrow HO_2^* \ [\text{-50/+120 mV (acidic, pH < 4.8)] (12)}$$

$$O_2 + e^- \rightarrow O_2^{*-} \ [\text{-330/-160 mV (neutral or alkaline)] (13)}$$

For the reduction of superoxide:

$$HO_2^* + H^+ + e^- \rightarrow H_2O_2 \ [\text{+1440 mV (acidic, pH < 4.8)] (14)}$$

$$O_2^{*-} + 2H^+ + e^- \rightarrow H_2O_2 \ [\text{+890 mV (neutral)] (15)}$$

$$O_2^{*-} + H_2O + e^- \rightarrow HO_2^- + OH^- \ [\text{+200 mV (alkaline, pH >11.5)] (16)}$$

As seen, protons play a critical role in several steps and affect both the thermodynamic and kinetic aspects involved. The production of superoxide is not dependant on protons, at least at the formation stage. Yet, the rate of superoxide availability would be critically dependant on protons because protonated superoxide (hydroperoxyl radical) is known to be many times more (re)active than superoxide itself. If the superoxide would get protonated near the membrane, it would be easily lost to dismutation in the matrix (leading to the liberation of heat energy). [For more details in the context, please refer Item 1, Supplementary Information.]

*Reallocation of roles:* Ubiquinone need not be seen as a two-electron agent alone, its one-electron reduced form is also quite stable and would serve to enhance radical lifetimes in the vicinity of the membrane. The presence of various metal centers on the different proteins (of varying redox potentials) is merely to have catalysts in various redox potential ranges so that the radical formation and release would be smoothly transitioned and consistently maintained. Therefore, it is now understood that the mitochondria primarily achieves a low-water, low-proton microenvironment. The outer phospholipid membrane forms the outer dome of the reactor and Cyt. *c* (functioning as a one-electron scavenger within the inter-membrane space) is the last barrier to prevent the escape of one-electron equivalents. Complex IV can accept the redox equivalents from Cyt. *c* and bring it back to the internal murzone, and also stabilize/reduce superoxide and further generate hydroxyl radical and singlet oxygen. Complex III could recycle the radicals/$CoQH_2$ that was formed in the internal membrane. Complex V serves chemostasis (proton server to the matrix) and it could have evolved later on, to enhance the efficiency of oxidative phosphorylation. Most functional roles are thus re-allotted.

*Understanding the kinetics, energetics and overall rationale:* The overall drive for the reaction comes from primarily a one-electron oxidation of suitable reducible substrate(s) and oxygen serves as the ultimate electron acceptor. But the electrons from NADH (or $FADH_2$) do not go to oxygen all the way via ETC to form water only at Complex IV. There are no specified loci for water formation and electrons need not be exchanged between adjacent centers alone. Elaborate scheme for the spatial separation of protons and electrons do not exist and are not warranted. However, protons must be made available only at key points, by enzyme activity, failing which uncoupling could result (leading to heat production). The generation of radicals drives both water and ATP formation in the low proton/water microcosm. The more water formation is associated with ATP synthesis; greater is the coupling in the system. The more water formation is owing to radicals reacting among themselves; lower is the coupling in the system. Stoichiometry in such a system would depend very critically on the reaction microenvironments and this could be a reason for the differences in data obtained across labs. The overall yield of ATPs would vary when starting from NADH or succinate under diverse molecular/microenvironment setups. The one-electron reactions are the mainstay of the actual fast processes that govern core of energy production of cellular life and the two electron reactions serve as electron sinks, generating pull to drive the reaction to the right. Once superoxide is produced, it goes via peroxide and hydroxyl radical to form water. With $FADH_2$, the excess protons could potentially limit superoxide availability, affording significant peroxide formation. The novel way of water formation is the most sensible scheme considering that these reactions would only be binary and very fast. As seen, the new proposal only requires an intact mitochondrial membrane and the minimal components of ETC for the *murburn* process to work. Most importantly, the *murburn* concept addresses the chemical, evolutionary, stochastic and physiological factors involved within the reaction system. Life evolved from chaos and the *murburn* hypothesis has chaos enshrined in its basic elements. Nature evolved by a controlled burning of redox equivalents (generated by the catabolism of nutrients), forming water at the phospholipid interface. Therefore, the erstwhile EPCR's presumed uncoupling reactions are what actually lead to "chemical coupling" of ATP synthesis. (This outcome is quite similar to the effect that murburn concept brought about in the mXM system.)

**3. Arguments/predictions validating murburn concept's relevance in mOxPhos:**

(i) The documentation of a naturally formed nuclear reactor on earth [Meshik et al, 2004] is testimony to the feasibility of spontaneous evolution and operation of the "nuclear reactor logic" of murburn concept for mOxPhos. (ii) The fact that DROS are generated "enzymatically" by Complexes I through IV (both *in vitro* and *in situ*) [Grivennikova & Vinogradov, 2006; Drose, 2013; Bleier & Drose, 2013; Ksenzenko et al, 1992] is direct evidence for murburn concept. (iii) Actively respiring and ATP synthesizing mitochondria are demonstrated to dynamically increase DROS production [Nicholls, 2004] and this is a direct and incontrovertible evidence to support the physiological relevance of murburn concept to ATP synthesis. Further, DROS production has also been correlated to the *in situ* concentration of oxygen and reduced substrates [Murphy, 1987]. (iv) A non-integral stoichiometry [Hinkle, 2005; Brand & Lehninger, 1977] is naturally expected in any system where murburn concept is relevant [Manoj et al, 2016d] and therefore, mOxPhos system is a classical example where it could be operative. (v) The organization of mitochondria with only cristae invaginations and two membranes (with the membrane proteins scattered therein) agrees well with murburn concept. (vi) The distribution of proteins shows low amounts of Complexes I & II (superoxide generators) and high amount of Complex IV (DROS stabilizer); and this is solicited by murburn concept. Further, the relative concentrations of Cyt. *c* and Complex IV also align well with murburn concept. [Gupte et al, 1984; Schwerzmann et al, 1986] (vii) The finger-like matrix projection structures of Complexes I & II [Sazanov, 2015; Sun et al, 2005] meet the requirements of murburn concept. The fact is also that they do not house redox centers with a perfectly increasing order of redox potentials. These are tell-tale signs that the redox centers produce and moderate DROS. (viii) The erstwhile consideration required the inner mitochondrial membranes to be discriminatory and regulatory in structure and function. The mitochondrial membranes are seen herein as a hydrophobic interface that could potentially support and confine radical reactions. For all practical purposes, mitochondrial membranes do not show "super-deterministic" ability/features. This ascertained fact aligns well to the mandate that murburn concept allocates them. (ix) Contrary to the aesthetic perceptions prevailing amongst several biologists and biochemists (that DROS are merely chaos infusing agents!), I have established over the last decade that *in situ*, a two-electron process can go through one-electron routes (without wreaking havoc!) and that such processes are essentially involved in routine redox metabolism [Manoj, 2006; Manoj & Hager, 2008; Manoj et al 2010a,b; Manoj et al, 2016a-d]. This fact can be confirmed with simple chemical controls and reductionist approach, within water-controlled systems (like reverse micelles [Han et al, 1987; Hayes & Gulari, 1990]) and using Fe/ peroxide/ NADH/ superoxide. (x) The fact that multiple $^{18}O$ atoms (from water, the solvent) get rapidly incorporated into a single ATP phosphate [Cohn, 1953] is a solid support for the murburn concept, which seeks precisely such effects. (xi) The heat generation by chemical uncouplers (interfacial DROS modulators) and uncoupling protein (which have positively charged amino acid residues within their trans-membrane helices that could modulate DROS [Klingenberg & Huang, 1999]) is yet another testimony to the operation of murburn concept. (xii) There is historical documentation (quite forgotten!) that uncouplers give concentration-dependent maverick effects in phosphoryating systems [Avron & Shavit, 1965; Watling-Payne & Selwyn, 1974], an effect which is predictable if murburn concept is operative. This could be probed again with diverse concentrations of known uncouplers and it would further establish murburn concept's physiological relevance. (xiii) Very high or very low concentrations of oxygen would inadvertently affect the subtle redox equilibriums in milieu (as predicted by murburn concept) and therefore, we observe oxidative stress in both scenarios. (xiv) Toxicity of low concentrations of agents like CO and cyanide can be better explained by murburn concept [Parashar et al, 2014], owing to their roles as diffusible radical generators and scavengers. This can be further verified by tracing the incorporation of radio-labeled carbon in carbonate and cyanate, after exposing an experimental system to sub-lethal amounts of radio-labeled CO and cyanide. (xv) The evolutionary implication of murburn concept's operation explains why leaching of Cyt. *c* leads to cellular apoptosis [Ow et al, 2008]. (xvi) Equations 7 through 16 could potentially explain how the provision of a proton gradient (when inside is alkaline) displaces $ADP + Pi + H^+ \leftrightarrow ATP + H_2O$ equilibrium and afford synthesis of ATP. With murburn concept, mitochondria can work even without a pH gradient, at low proton concentrations and this goes well with several experimental observations. (xvii) Murburn concept can explain the influence of membrane lipids like cardiolipin [Paradies at al, 2014], whose high negative charge densities on their lipid-assembled periphery would keep superoxide in the matrix side of the interface, thereby enhancing phosphorylation yields. Murburn concept could also explain the toxicity of cationic lipids used for drug delivery [Aramaki et al, 2001]. (xviii) The murburn scheme is energetically sensible, viable and could potentially afford higher efficiency for ATP synthesis. (xix) Cytochrome *c* and CoQ have also been allocated one-electron scavenging/recycling roes within the IMS and IPLM respectively. Their concentrations, structure and nature agrees with the roles allotted. (xx) Murburn concept does not seek a vitally deterministic scheme and does not necessitate "overt intelligence" (sensory and regulatory discriminations in redox activities) to molecules. The proposed mechanism is simple (supported by Ockam's razor), unordered (no sequential schemes and no multi-molecular reactions sought), highly probable, fast (with sufficient thermodynamic drives and kinetic viability- only radical reactions mediated by small species could give second order rates ~$10^9$ to $10^{10}$ $M^{-1}s^{-1}$) and thus, can potentially cater to the dynamic requisites of a working cell. [For a more detailed discussion, please refer Items 2 & 3, Supplementary Information.]

**4. Consequences of this communication:** An itemized comparison of the prevailing EPCR hypothesis versus the proposed murburn concept is provided in Item 4, Supplementary Information. While the new explanation undoes several misunderstandings, it opens up a new bunch of questions- In such murburn processes, does water break, besides forming? What is the concept of "phosphorylation site"? Do the Complexes I through IV have any ADP binding sites? Does ATPase bring protons in along the b subunits to help the coupling process at one "new" site (adjacent to the region where b-α/β/δ subunits meet)? What is the role of the bulbous matrix-side extension of Complex III? Is ATPase actually a rotary enzyme? etc.

In mOxPhos, biochemists erroneously looked at the cause-consequence correlation. Murburn concept puts things in the accurate directional perspective. A trans-membrane potential is generated because of NADH oxidation inside (leading to negative charge-bearing ionic- and radical species generation within matrix) and not because protons are pumped out. This has little to do with synthesis of ATP inside. Further, the existing explanations solicited highly sophisticated setups (ETC, proton pumps, rotary synthases, etc.; all of which were debunked [Manoj, 2017]) as a pre-requisite for the evolution of the energy transduction logic/process in living systems. Herein, we have the first molecular explanation for the origin of life-sustaining "tangible" logic of chemical reaction. This could also explain why only a pre-existing cell could infuse "life". The radical reaction requires an internally "burning flame containing mitochondria" to start off, and only a pre-existent cell could pass it on to a new cell. Cells burn fuel and in this process, they too get burnt and as a result, they age and they die (oxygen, the life-giver, finally takes it too!). But I believe that with the current awareness, ageing and death can be better understood and in some cases, even death of some cells could be rendered reversible. I envisage that murburn concept would be applicable to a wide variety of physiological processes that involve redox activities and response to physical stimuli (inputs like pressure, exposure to light, etc.). It is now evident that murburn concept can also provide a molecular basis for idiosyncratic and hormetic dose responses. The work calls for a renewed investigation into all redox coupling reactions where an uphill reaction is catalyzed at the expense of a downhill reaction. We must derive tangible rationale as to how "energy" (a term coined by physicists to rationalize and account for relatively abstract ideas/events) is "t(r)apped" and "relayed" / "transduced" from one system to the other.

**TABLE & FIGURE**

**Table 1:** Similarities and parallels (as hitherto perceived!) between microsomal/hepatocyte ER xenobiotic metabolism and mitochondrial oxidative phosphorylation. Many of the mechanistic perceptions in the microsomal system have been corrected; they are presented herein only to draw an existential analogy.

| No | Attributes | | Microsomal system | Mitochondrial system |
|---|---|---|---|---|
| 1 | *Various facets of the reaction system* | Overall reaction | Redox reaction + addition of moiety | Redox reaction + addition of moiety |
| 2 | | Initial electron donor | NADPH | NADH |
| 3 | | Final electron acceptor | Molecular oxygen ($O_2$) | Molecular oxygen ($O_2$) |
| 4 | | Obligatory Catalyst 1 | Reductase (flavoprotein CPR, binds and uses NADPH); membrane protein | Dehydrogenase (flavoprotein Complex I, binds and uses NADH); membrane protein |
| 5 | | Obligatory Catalyst 2 | Mixed function oxidase (hemoprotein CYP, binds and uses oxygen); membrane protein | Oxidase (hemoprotein, Complex IV, binds and uses oxygen); membrane protein |
| 6 | | Ratio of catalysts | Low flavo protein : high Hemo protein | Low flavo protein : high Hemo protein |
| 7 | | Reaction aids | Cyt. $b_5$ (& Fe-S proteins in auxiliary systems) | Cyt. c, Fe-S proteins & CoQ |
| 8 | | Reaction centre-stage (in vivo/in vitro) | Phospholipid interface/vesicles | Phospholipid interface/vesicles |
| 9 | | Final product of interest | Xenobiotic is hydroxylated | ADP is phosphorylated |
| 10 | | System overview | Multi-enzymatic ordered sequential scheme, multi-substrate, multi-product, coupling/uncoupling phenomena | Multi-enzymatic ordered sequential scheme, multi-substrate, multi-product, coupling/uncoupling phenomena |
| 11 | *Reaction phenomena & mechanism* | Partitioning of substrate and product at the membrane | The hydroxylated product has a lesser log P value than the substrate. | Log P value of ADP is -2.6 and that of ATP is -5.5. |
| 12 | | Electron transfer | Long range, outer sphere mechanism; Essentially two-electron process overall, could involve one- | Long range, outer sphere mechanism; Essentially two-electron process overall, could involve one- |

| | | | electron step(s) | electron step(s) |
|---|---|---|---|---|
| **13** | | Protons | Extraneous proton consumed | Extraneous proton consumed |
| **14** | | Thermodynamic drive | Push of electrons from lower to higher redox potential, through direct protein-protein contact | Push of electrons from lower to higher redox potential, through direct protein-protein contact |
| **15** | | Substrate binding | Xenobiotic binds to catalyst 2 for prolonged time frame and waits for catalytic cycle to get completed | Oxygen binds to catalyst 2 for prolonged time frame and waits for catalytic cycle to get completed. |
| **16** | | Water formation | Catalyst 2 forms water | Catalyst 2 forms water |
| **17** | | ROS involvement | ROS formed as a deleterious side reaction, due to uncoupling | ROS formed as a deleterious side reaction, due to uncoupling |
| **18** | | Inhibitors like CO and CN$^-$ | Bind to Fe-centre of catalyst 2 | Bind to Fe-centre of catalyst 2 |
| **19** | | Salient uncouplers | Dihalophenolics | Dinitrophenolics |
| **20** | | Overall stoichiometry | Subject to change by alteration of reaction microenvironment and non-integral values of ratios noted | Subject to change by alteration of reaction microenvironment and non-integral values of ratios noted |

Figure 1: **a.** Cross-sectional view of the mitochondrial membranes to show the relative distribution and rough dimensions of various complexes' penetration into the matrix in murzone. (Figure is not drawn to scale, but for deriving semi-quantitative impressions alone.) The ratio depicted is a I:II:III:IV:V:Cyt. $c$ = 1:2:2:6:6:9, based on the subjective average of Gupte et al [1984] & Schwerzmann et al [1986] data. The reaction system can be considered analogous to a nuclear fission reactor, the difference being that DROS are involved herein. The putative enzymatic sources of superoxide are drawn in bold. Complex III's role needs to be redefined/recharted. It can be seen that the erstwhile EPCR hypothesis has little scope for any effective functioning with such a distribution. For details, please refer Item 1, Supplementary Information. **b.** Surface view depicting relative distribution densities of the complexes on the inner membrane [Gupte et al, 1984; Schwerzmann et al, 1986]. The images are not drawn to give precise spatial arrangements, but only to indicate an overall distribution. Depiction key:  Complex I - circle, Complex II - plus star, Complex III - triangle, Complex IV - rhombus, & Complex V - pentagon, Cytochrome $c$ - asterisk. Cytochrome $c$ is not shown in the overall view (to retain clarity in presentation). In the individual systems shown on the right, Gupte's values are weighted.

**a.**

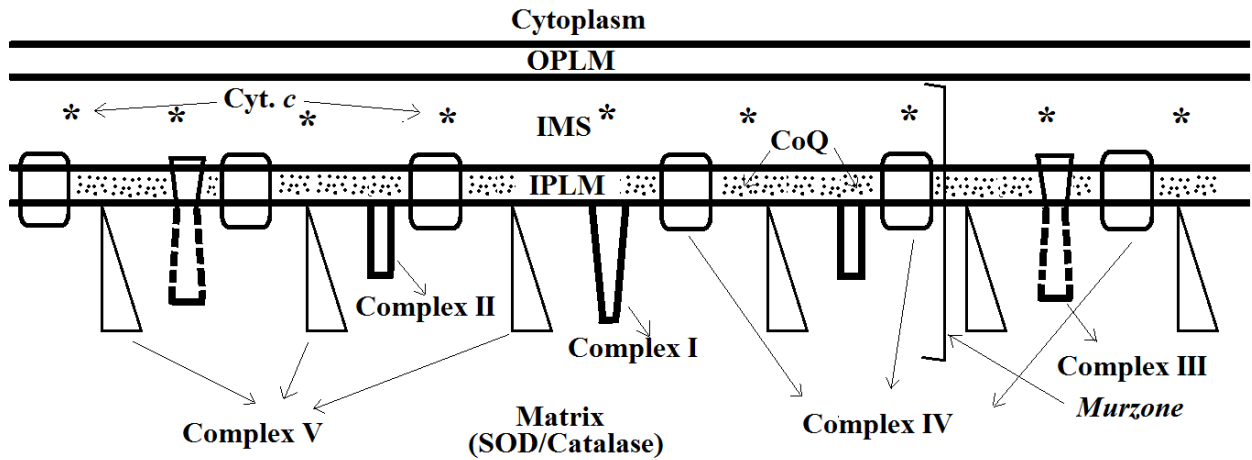

**b.**

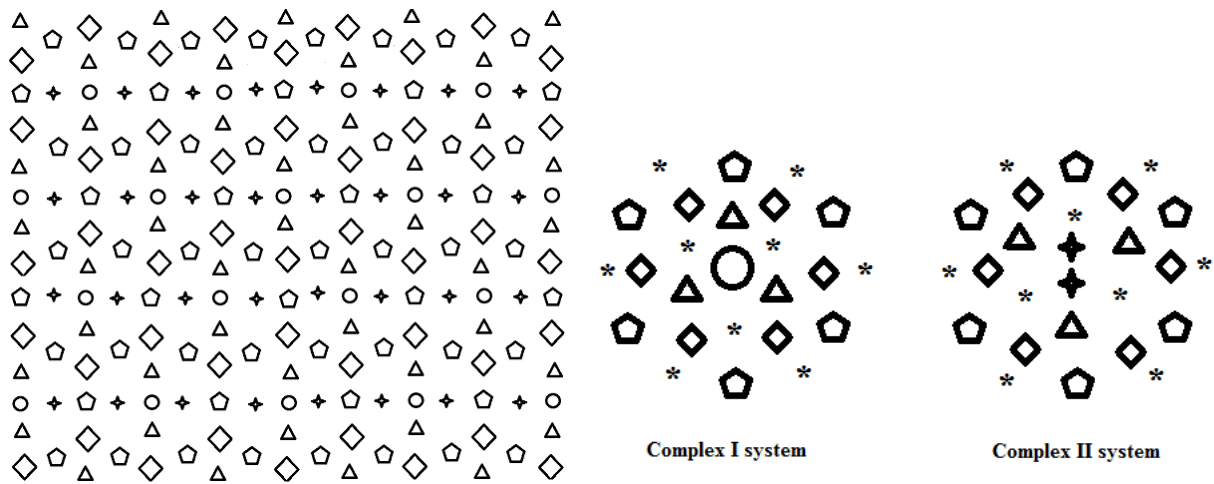

# Supplementary Information: *Murburn concept…………*

## Kelath Murali Manoj (2017)

### *Materials & Methods*

All data I have perused are other researchers' hard work. While attempting to unravel the puzzle of mOx-Phos, here is what I have tried to adhere to-

1. Start with doubts!

2. Pursue "doubt clearing" agenda starting from known facts, framing appropriate/right questions.

3. Challenge all protocols/inferences that involve undue complexities and interpretations that are based on "unfounded aesthetics or preconceived notions".

4. Prioritize on putting in the bigger pieces of the puzzle in first as nucleation points. Since we deal with reactions, always prioritize space-time constraints and balance mass-charge movements. [Using these procedures, the hitherto prevailing explanations have already been debunked.]

5. Form new hypotheses only if there is a need and adequate information for the same. Otherwise, stay on exploratory mode.

6. Use Ockham's razor and evolutionary perspectives as guides to piece in the remaining puzzle.

7. If ambiguities and uncertainties linger, acknowledge them and leave room for revamping and refinement.

8. End with doubts!

## Item 1: Details of Murburn concept with mOxPhos perspective

(Before reading the following text, it is strongly advised that the reader peruse the discussion on "debunking ETC concept" and peruse the relevant citations in that manuscript.)

*Initiation:* Let's take the Complex I - centered system alone first. NADH could give one or two electrons to this complex. (Please see that direct evidence of NADH radical formation is available [Zielonka et al., 2003]. Besides, one can envisage that an attack by an *in situ* formed hydroxyl or perhydroxyl radical could also leave behind such a radical.) Regardless of whether NADH gives one or two electrons to the flavin of Complex I, the flavin does not have a choice to pass on two electrons down the stalk, all the way to CoQ (the erstwhile purported acceptor). Therefore, oxygen could receive at least one electron and establishes a one-electron paradigm within the milieu immediately surrounding Complex I. The two redox centres (N1a and N3) located adjacent to the flavin can help this process by taking the two electrons from flavin (one apiece or just one at a time!) and then relaying one down the stalk and relaying the other to oxygen. Oxygen can collect the electron directly from the flavin also, though it would be a relative slow process for the first step. But once the "chain reaction" sets in, scenario changes, as the enzyme could recycle via the semiquinone step or singlet oxygen could be generated from the other metal centres in the murzone. The two "non-reducible centres" in the stalk would afford ample scope for generating superoxide by ensuring that a one-electron paradigm sets in owing to electron abstraction from NADH. (This could be due to their inability to receive electrons from the "chain". Else, if it did receive the electrons, they could easily reduce molecular oxygen on their own, owing to their low potentials!) Let's remember that mitochondrial matrix is a low-proton environment and therefore, superoxide should be relatively stable in this regime (to have at least a couple of 1-5 nm motility range). The semiquinone formed at the flavin would earnestly give one-more electron to the next oxygen. The quicker electron relay/reactions are the single electron reactions. Unaided one-electron transfers between redox centers can occur if the potentials are favorable and the distance is small. Otherwise, help has to come from diffusible one-electron equivalents (primarily, superoxide) generated in situ. Now, superoxide (not a caged-radical bound to flavin anymore!) is now free to move around the murzone, reducing several reaction centres and redox-active molecules within the system (depending on the ambient concentration of superoxide itself, and on the concentration of NADH and oxygen). A given Complex within the reaction system could thus need to have different redox centres of diverse redox potentials, to ensure a give and take of one-electron equivalents across various regimes and dynamics of oxygen-ROS. (This is the moderated release of one-electron equivalents.)

Now, let us get to Complex II. Here, the flavins are at a much higher redox potential and it is highly unlikely that oxygen can receive an electron at this step. Besides, the donor/acceptor is a tightly bound $FADH_2$ with protons being "satisfied" (unlike the case of NADH, where a proton was "short"!). But the transfer from the first Fe-S (2Fe-2S) to the second Fe-S (4Fe-4S) centre is rather unfavorable (both in terms of both distance and redox potentials) and therefore, oxygen still has a fair chance to grab the electron. (This would result if superoxide concentration is low and if oxygen concentration is high.) Therefore, ROS generation from this centre would not be as efficient as it is at Complex I. (But that would not matter much because electrons put into this system from multiple points have a very high probability of getting back into the murzone, which can extend up to 150 Å into the matrix, from the inner mitochondrial membrane. Kreb's cycle would lead through succinate-fumarate-malate, the last of which would anyway generate NADH. It must be noted that the starting points of both "ETCs" are flavoproteins, well-known for their superoxide generation ability. Further, the two other electron donation portals into the mitochondrial system, glycerolphospate dehydrogenase and acylCoA dehydrogenase, are also flavoenzymes! But it is notable that these enzymes do not possess long-finger like projections containing multiple metallic redox centers [Yeh et al, 2008; McAndrew et al, 2008].

There is a catch to the initiation-sustenance process- If there is no "slow 2-electron sink" available immediately nearby, the initiation process of oxygen-superoxide equilibrium does not move to the right (especially, through CoQ reduction). This is why Complexes I & II are inhibited by molecules like rotenone and carboxin (respectively). Once any Complex is inhibited by such molecules (as further exemplified by stigmatellin and antimycin A for Complex III), we can see the relatively slow accumulation of some stable reduced redox centers upstream (their reduction sponsored by the redox equivalents that remain). This is the direct evidence that the system is geared to retaining a one-electron reservoir within. Each complex has the ability to sustain a murzone around itself.

*Sustenance:* Now, the electron charged or deficient species would have a slow propensity to traverse toward the membrane because of the redox centres closer to the membranes have higher redox potentials. (Also, the phospholipid membranes, with their hydrogenated carbons would have higher electron densities, compared to the aqueous environment in the matrix. Thereby, they would "invite" the electron deficient radicals their way.) It is highly probable that the superoxide would meet the highly concentrated CoQ in the inner membrane. By chance, if it goes through the membrane without meeting a CoQ or another redox centre of another system, it would be picked up by Cyt c at the IMS, only to be relayed back into the system via Complex IV. If the CoQH formed does not meet an oxygen molecule quickly enough, $CoQH_2$ may be formed, which would be acted upon by Complex III. (But the latter types of two-electron reactions would be relatively on the slower time scale.)

Complex III could give one electron to Cyt. c on the IMS side. (But I think that the large bulbous extension towards the matrix side must have some hitherto undiscovered functionality. It could be an oxygen or DROS conduit, to channelize at least one electron back into the matrix side. Thus, Complex III could also re-establish a one-electron paradigm and recycle lost electrons! It could be a membrane bound peroxidase or even an ADP binding protein too. Now that the complex CoQ cycle is debunked and deemed highly unlikely, the functionality of this enzyme needs to be reinvestigated in minimalist setups, with a fresh approach, to ascertain its role. The respirasome structure gives it a strong membrane anchoring/clipping role, with respect to Complex I. This makes sense with murburn concept because we cannot afford Complex I to get uprooted from the membrane at any cost; because in a soluble state, it would sponsor uncontrollable chaos.) It is not necessary that any two-electron stabilized intermediate (in milieu) should receive both of its electrons from a given/destined provider alone. All participants within the murzone would function via a non-specific one-electron scheme (barring select reactions like the electron withdrawal step from succinate dehydrogenase). If one-electron removal from NADH occurs, then the NADH radical formed could also function as an electron (cum proton!) donor within the murzone.

If lots of Cyt. *c* got reduced, it means that the system is getting overloaded and therefore, the reaction is going too high. (This would be quite unlike the requirement of the earlier explanation, which would need one-electron species to recycle via the IMS, obligatorily through Cyt. *c*, at high fluxes!) Herein, Cyt. *c* is seen as the penultimate agent (the last suicidal barrier being the OPLM!) in the system that traps one-electron equivalents. As per the textbook value [Lehninger, 1990], a two-electron redox potential of +45 mV is given for the $CoQ$-$CoQH_2$ redox couple. As the pH becomes more alkaline, CoQ becomes more difficult to reduce, with a $\Delta E/\Delta pH$ value of -60 mV [Urban & Klingenberg, 1969]. It is interesting to note that CoQ's one-electron redox couples have been attributed (determined or calculated or projected) a wide range of values, from

-300 mV all the way to +200 mV, for the various redox couples like- $CoQ$-$CoQ^{*-}$, $CoQ$-$CoQH$, $CoQ^{*-}$-$CoQH_2$ and $CoQH$-$CoQH_2$ [Kindly refer the tables in my "Debunking ETC" manuscript and Roberson et al, 1984; Verkhovskaya et al, 2008; Krumova & Cosa, 2016; etc.]. Therefore, we can now truly appreciate why CoQ is present within the membrane- it serves as a redox buffer (both one electron and two electron!) in a wide range..

Now, any Fe complex with a penta-ligated heme (like Complex IV) could bind superoxide, thereby stabilizing it. Complex IV qualifies for the role. It can stabilize radicals (superoxide, perhydroxyl & hydroxyl) and hydroperoxide ions. And it can also generate singlet oxygen, hydroxyl radical and peroxide in milieu. All these species have a finite lifetime around the inner membrane of the matrix. If Complex IV receives an electron while superoxide is bound, peroxide ion could be generated. Now, we have achieved the precondition for Fenton chemistry within the murzone. Sponsored by NADH, Fe centres can carry out Fenton chemistry within the potential range of ~-320 mV to ~+460 mV (the latter value being the midpoint for $H_2O_2 + e^- \rightarrow OH^- + OH^*$). The window for the same effect with superoxide would be above -160 mV (at equimolar oxygen to superoxide ratio). Say, if the superoxide concentration is only a thousandth or millionth of the oxygen concentration, then only iron centres with +23 mV or +206 mV respective redox potentials can qualify for reduction in this scenario. (Now, we can see why succinate affords lesser ATP than NADH. The earlier explanation that there is little potential drop between Complex II – CoQ / Complex III electron transfer does not make sense because one of redox centers of Complex II has a midpoint potential of -260 mV.) Therefore, Complex IV is a key agent to keep things "cool" (binding and releasing the radical in a sustained manner) and "heat" things up too (because its activity could generate the high potential hydroxyl radical). That is why there are 6 of them per Complex I. Now, any peroxide or superoxide that leaches out into the matrix or cytoplasm is scavenged by catalase and superoxide dismutase (SOD). Enzymes like GSH take care of the lipid peroxidations that result in the process. Under high productivity, if a hydroxyl radical meets CoQH or reduced Cyt. *c* on the way out into the cytoplasm, the latter happily imparts an electron to this highly reactive agent, reducing the radical to hydroxide ion. Now, the low redox potential centre of Complex II (4Fe-4S, whose presence meets no explanation in the erstwhile perceptions!) would have a purpose- under high electron flux, it could get reduced and release the same later.

Negatively charged ions are formed due to the initiation and progression of NADH/succinate oxidation in the murzone. To counter the same, if protons move in passively, then energy conservation in ATP formation is efficient. If the rate of respiration is very high, then diffusion alone cannot counter for the proton deficiency. This is because resistance of the membrane cannot change temporally owing to a vital deterministic need. Resistance of the membrane is an important feature for maintaining the respiration itself. The phospholipids would ensure that much of the negatively charged superoxide does not traverse the membrane; and cardiolipin's presence in mitochondria would be very efficient at ensuring that outcome. The presence of a greater negative charge density would repel superoxide's approach to the membrane, keeping a higher concentration of the same within the matrix side of the murzone.

Now, the uncatalyzed radical reactions in murzone would be [Sawyer & Valentine, 1981; Buxton et al, 1988; Gutowski & Kowalszyk, 2013]-

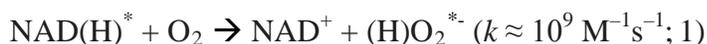

$NAD(H)^* + O_2 \rightarrow NAD^+ + (H)O_2^{*-}$ ($k \approx 10^9\ M^{-1}s^{-1}$; 1)

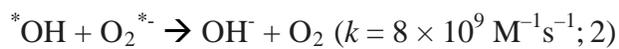

$^*OH + O_2^{*-} \rightarrow OH^- + O_2$ ($k = 8 \times 10^9\ M^{-1}s^{-1}$; 2)

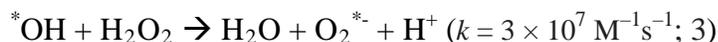

$^*OH + H_2O_2 \rightarrow H_2O + O_2^{*-} + H^+$ ($k = 3 \times 10^7\ M^{-1}s^{-1}$; 3)

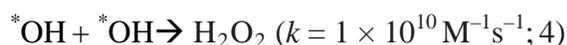

$^*OH + {}^*OH \rightarrow H_2O_2$ ($k = 1 \times 10^{10}\ M^{-1}s^{-1}$; 4)

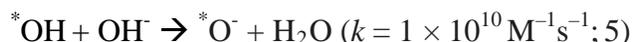

$^*OH + OH^- \rightarrow {}^*O^- + H_2O$ ($k = 1 \times 10^{10}\ M^{-1}s^{-1}$; 5)

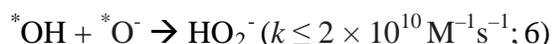

$^*OH + {}^*O^- \rightarrow HO_2^-$ ($k \leq 2 \times 10^{10}\ M^{-1}s^{-1}$; 6)

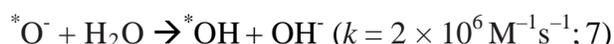

$^*O^- + H_2O \rightarrow {}^*OH + OH^-$ ($k = 2 \times 10^6\ M^{-1}s^{-1}$; 7)

As seen, the reaction microcosm would comprise of multiple one- and two- electron redox active species and equilibriums; and flavin/heme/Cu catalysts would entertain many types of inter-conversions, making the reactions thermodynamically and kinetically feasible/viable (though apparently chaotic!). The one-electron redox exchange systems are relatively less energetically expensive and thus, more reversible. Some of these reactions might lead to loss of redox equivalents (or quench the ROS/radicals formed; particularly when two dissimilar radicals collide) or keep the highly fast radical reactions going. Since we are dealing with physiological pH, with a few tens of protons, we can safely discount many reactions that seek protons' active involvement. Reaction schemes involving three distinct agents coming together are not considered probable. Further, one-electron reactions would be favored because there are a lot of one-electron moderators around. That leaves behind only a few uncatalyzed reactions within the milieu. We see that superoxide can be stabilized within the murzone, without causing major chaos. Superoxide's reduction to peroxide and its subsequent reduction to hydroxyl radical could be catalyzed by Complex IV or this could happen around its vicinity. Complex IV has several trans-membrane helices that could fetch the proton needed for this activity from the outside. (I doubt if Complex II/III's heme matrix-ward heme be involved in this step because both are hexaligated.) So, one can gauge that the reaction milieu is not that chaotic, after all! (Since the reduction potential of unsaturated fatty acyl chains is around +600 mV, both perhydroxyl and hydroxyl radical are capable of wreaking havoc. Therefore, these radicals' production must be controlled and minimized or they must find another suitable reactant quite close to their generation point.) The DROS concentrations/ lifetimes are controlled in the murzone by the finger projections of Complexes I & II, CoQ / Cyt. *c* and their collisions are moderated by the intrinsic reaction-generated hydroxide ions ($OH* + OH^- \rightarrow OH^- + OH*$). CoQ scavenges radicals within the membrane and Cyt. *c* latches on to any one-electron species that manages to escape into IMS. The essence of this idea would be clear if one sees the density distribution of the pertinent proteins / small molecules and the cross-sectional view of the reaction system (Figure 1a and 1b of the main article).

*The coupling step (termination):* As can be gauged, the formation of water is the ultimate termination step. At the physiological pH of 7.4, inorganic phosphate (Pi) has a net negative charge and exists more as $HPO_4^{2-}$ & less as $H_2PO_4^-$. This is because its $pK_a$ falls slightly below neutral pH at physiological temperatures. Further, ADP & ATP (the adenosine base being represented as "Ad" below) also have a net negative charge (at the physiological pH of 7.4), with comparable $pK_a$ values (in the pertinent range) as that of Pi. [[The values that I cite are from Paula Bruice's textbook- Organic Chemistry. They are- ATP: 0.9, 1.5, 2.3 & 7.7 with a net charge of -3.3; ADP: 0.9, 2.8 & 6.8 with a net charge of -2.8 and Pi: 1.9, 6.7 with a net charge of -1.8. Several sources put the pertinent reactive species' $pK_a$ to be in the 6.7 to 7.7 pH range. Herein, I have not considered the complexities brought in by the complexation with magnesium ions, which exists in realistic physiological states. The magnesium concentration is usually high (in the range of a few tens of mM), and the matrix pH supposedly remains above 7.2.]] Therefore, the following equilibrium equations then become relevant:

$H_2PO_4^- + H_2O \leftrightarrow HPO_4^{2-} + H_3O^+$ (8)

Ad-O-$(PO_2)^-$-O-$(PO_2)^-$-OH + $H_2O$ ↔ Ad-O-$(PO_2)^-$-O-$(PO_3)^{2-}$ + $H_3O^+$ (9)

Ad-O-$(PO_2)^-$-O-$(PO_2)^-$-O-$(PO_2)^-$-OH + $H_2O$ ↔ Ad-O-$(PO_2)^-$-O-$(PO_2)^-$-O-$(PO_3)^{2-}$ + $H_3O^+$ (10)

Now, in these pH ranges, the phosphorylation reaction could be as per the following schema-

Ad-O-$(PO_2)^-$-O-$(PO_2)^-$-OH + $H_2PO_4^-$ ↔ Ad-O-$(PO_2)^-$-O-$(PO_2)^-$-O-$(PO_2)^-$-OH + $H_2O$ (low pH; 11)

Ad-O-$(PO_2)^-$-O-$(PO_3)^{2-}$ + $H_2PO_4^-$ ↔ Ad-O-$(PO_2)^-$-O-$(PO_2)^-$-O-$(PO_3)^{2-}$ + $H_2O$ (~mid/neutral pH; 12)

Ad-O-$(PO_2)^-$-O-$(PO_2)^-$-OH + $HPO_4^{2-}$ ↔ Ad-O-$(PO_2)^-$-O-$(PO_2)^-$-O-$(PO_3)^{2-}$ + $H_2O$ (~mid/neutral pH; 13)

Ad-O-$(PO_2)^-$-O-$(PO_3)^{2-}$ + $HPO_4^{2-}$ + $H^+$ ↔ Ad-O-$(PO_2)^-$-O-$(PO_2)^-$-O-$(PO_3)^{2-}$ + $H_2O$ (high pH; 14)

Mitochondria work when internal pH is ~7.3 or slightly higher. We can envisage that lowered similar-charge repulsions facilitate better effective collisions/orientation of the reactants (ADP & Pi) and the presence of the "dissociable proton" enables a water molecule to "leave" from the penta-covalent phosphorus intermediate (in case of a nucleophilic attack), thereby leading to efficient ATP formation [Stryer, 1990]. (Therefore, higher pH would be detrimental to ATP synthesis.) Mechanistically, the same logic would hold if the reaction were to be facilitated by a diffusible species (like hydroxyl radical mediating an electrophilic attack) in milieu. A hydroxyl radical could attack the highly abundant phosphate ion thereby activating it, which in turn, can add on to an ADP molecule to give ATP. (Not too far-fetched, when it is known that hydroxyl radical reacts with oxidized ions like carbonate, at rates approaching 4 x $10^8$ $M^{-1}s^{-1}$ [Buxton et al, 1988]!) Whatever the "actual route", it is logical to assume that with tens of millimolar levels of inorganic phosphate around, it would chance to be the first scavenger of the low amounts of hydroxyl radical that would be formed. With this supposition, we have a tangible chemical connectivity/coupling mechanism between the oxidation of NADH and synthesis of ATP. This mechanism could also explain the high amount of radio-labeled solvent oxygen atom insertions in the ATP phosphate formed [Cohn, 1953] and also the inability of researchers to isolate a high-energy enzyme-phosphoryl intermediate. The pseudo-first order rate of ATP synthesis (~$10^3$ $s^{-1}$) can most probably be achieved only by a radical mediated process (which are known to give second order values of $10^9$ – $10^{10}$ $M^{-1}s^{-1}$). (On the other hand, a reductive addition of two electrons into the phosphate could lead to an oxide ion abstraction, to give a transient phosphite, which can get coupled with ADP. This suggestion was originally made by Mitchell [1985]. But for this scenario, the potentials have to be too low. However, the requirement of proton in the overall equation at alkaline pH also suggests a catalytic site mediate process. Therefore, overall mechanism needs to be investigated to ascertain the exact route.)

The chemical coupling logic of the reaction system would then be that most water molecule formation within the milieu should be via ATP synthesis. The more water is formed without radicals getting ample opportunities to interact with Pi, lower is the yield of ATP! At this juncture, the exact location of ATP formation is not nailed. It is not clear to me how Complex V can present ADP and Pi effectively on its surface/crypt to be catalytically coupled by the radicals, going through a rotary synthetic mode. Therefore, by all probability, the initial/primary source of ATP generation could be free radical mediated coupling within the murzone, in the matrix side of the inner membrane. If Complex V indeed serves as the locus to bind ADP and synthesize ATP, it cannot be owing to a rotary modality (as this can only be supported for hydrolysis function!). As a deduction, then the loci for ATP binding (for hydrolase activity) must be distinct from the loci for ADP binding (for esterification activity). (In this context, it would be worth investigating if all the other Complexes have any ADP binding site(s). I keep wondering why Complex III has the bulbous projection towards the matrix and what is the significance of the "three phosphorylation sites"?! If Complexes I, III & IV had ADP binding sites, then, we can reason why blocking these protein complexes inhibits ATP synthesis. As of now, since these three Complexes are involved in generating key intermediates and sustainment of a one-electron paradigm in milieu on their own merit, the relevance of murburn concept is justified.)

It can be seen from equations (8-14) that the $pK_a$ values of ADP and Pi are important; at pH higher than its $pK_a$, it would need to go through the reaction 14. That is- in the formation of a bond between ADP and Pi, the mono-negative Pi (the form on the left of equation 8) and/or the relevant protonated form of ADP could be the preferred reactant(s). It is known that free energy term for ATP synthesis goes up with increase in pH; that is- a higher numerically negative value is obtained with increasing pH. In other words, as the pH increases (hydroxide ion concentration goes up!), ATP hydrolysis becomes more and more favorable. Hydroxide ions are known catalysts that aid in a quick attainment of the ester hydrolysis equilibrium (to the extreme right, giving fully hydrolyzed products). Since NADH oxidation leads to the formation of hydroxide ions, it would necessitate a proton delivery system within the milieu (for an effective continuation of the reaction scheme). Therefore, to neutralize the hydroxyl ions (to meet the proton deficit), protons could move spontaneously inward through the membranes. If the rate of substrate oxidation is higher than the rates of proton permeability across the membrane, then it would be natural to hypothesize that ATPase could have evolved to aid the proton movement (internal proton deficit). It could achieve this mandate by hydrolyzing ATP, which would liberate a proton in the inside and also bring in 3-4 protons from the outside! This would also keep the ATP concentration low within the matrix (albeit at the cost of efficiency!), further aiding the equilibrium for ATP synthesis. (One molecule of ATP hydrolyzed is associated with the availability of 4-5 protons; which can be used to synthesize that many more ATP. Therefore, it is not a bad deal!) This supposition would explain why Complex V has a high affinity for ATP. The maintenance of pH in matrix is obligated to ensure the optimal nature of the concerned reactants and serve to minimize the competitive reactions that could otherwise set in. In such a reaction scheme, the pH would not only be important for ATP synthesis, but also for governing the lifetimes and interactions involving the radical and reactive oxygen species generated in milieu. [[Therefore, Complex V must make the proton available to the reaction milieu without messing up the superoxide dynamics. If Complex V indeed served as the "ADP + Pi coupling unit", the afore-mentioned outcome could be achieved with the two long b subunits serving as a proton conduit from the membrane to the synthesis site. With this perspective, Complex V does look to be a suitable coupling locus! (But it must also be borne in mind that this modality cannot work continuously in the rotary synthesis mode.)]] Overall, the system appears to have evolved to match the production of hydroxide ion (and hydroxyl radical with which it exists in equilibrium, in the electron-rich ambiance) by supplementing protons' inward movement with ATPase activity. [[The erstwhile view is that protons' mechanical movement inward leads to the synthesis of ATP at Complex V. Herein, I have posited that the synthesis of ATP is chemically (not just energetically!) coupled with the NADH oxidation! Oxidation of NADH leads to DROS generation and the in situ generated DROS serve as coupling catalysts. Further, I have proposed that Complex V hydrolyzes ATP to bring protons inward so that reaction dynamics is retained, albeit at the expense of some energy. It has been argued beyond reasonable doubt through my earlier manuscript that "Complex V cannot be a perfectly reversible enzyme.]] Thus, arguments have been presented that lead us to conclude that Complex V could serve to indirectly couple NADH oxidation to ATP synthesis. Now, we can also understand why addition of extraneous protons yields ATP! (It is because protons are reactants, which affect rates of several reactions and affect multiple redox equilibriums in the system. It is not because they serve towards any proton motive force. The equations presented in the main manuscript should serve to convince the reader of the veracity of this deduction.) Further, the system's basic conduction is self-regulatory and does not need external intelligence or sensors to keep it going.

I posit that the low water activity microenvironment within matrix further serves to enhance lifetimes of the catalytic radicals and thermodynamically drive the phosphorylation reaction forward. This is particularly of high relevance because the highly involuted mitochondrial matrix (with cristae invaginations) would be a low-water-water activity micro/nano-cosm. Quite akin to the low proton concentration found in mitochondria, another facet that requires great attention is this apparent low availability of water. If we go purely by the number of water molecules that 0.2 $\mu m^3$ volume of 55.6 M solution of water (which is what we assume for the calculation of energetics involved!) gives, we would have 6.6 x $10^9$ molecules of water. (Milo and Phillips [2015] calculate that the number of water molecules in a bacterium of similar dimensions would run down to 2 x $10^{10}$.) But the available volume must also account for the cristae invaginations and house the diverse proteins, metabolites and inorganics; and all this would consume space. A significant portion of the water present within the mitochondrion would not be available as "free water", because much would go into hydration shells of inorganics/ions, metabolites and proteins, etc. Coupled with the "matrix to cytoplasm" turgor movement of water through aquaporins and working of ATP-ADP antiporters, the reaction environment would effectively tilt the equilibrium (ADP + Pi $\leftrightarrow$ ATP + $H_2O$) towards the synthesis of ATP.

ATP is not stable in non-buffered aqueous solutions because its hydrolysis is thermodynamically favored. (This is because- 1. The bonds of hydration for ADP and Pi are stronger than the phospho-anhydride bonds of ATP. & 2. The resonance stabilization of products is more efficient and the charge densities are lower on the products than on ATP. Therefore, any aqueous solution of ATP would ultimately dissociate completely to ADP with the passage of adequate time. *So,* *we can see that ATP's bonds are not really that strong, after all!*) The actual reason that ATP hydrolysis affords a means of doing chemical work is because of the fact that the in situ concentration of ATP is significantly higher to that of ADP. (We know that any system that is displaced from its equilibrium state can do work!) As a consequence, the energy needed to synthesize or breakdown ATP is not fixed but varies on a diverse set of parameters. It is very intriguing to note that though the ATP formation/hydrolysis equilibrium is reversible, the ATPase cannot function reversibly as a hydrolyzer and synthesizer of the same molecule, via the same sites/mechanism. We know that practically, the facet of reversibility is an engineering feat very difficult to achieve in real space and time for a rotary trans-membrane enzyme. ATPase should be seen as a unidirectional enzyme that aids ATP synthesis in an indirect way. This conjecture is solidly supported by the fact that ATP has a $10^7$ folds higher affinity for ATPase. (If it was indeed a synthase, it should have had greater affinity for ADP!). It could have been appended to the respiratory machinery in later times, to meet the greater temporal demands of the cell. The original synthesis of ATP must be mediated via the radicals in situ, within the murzone. Therefore, the murburn hypothesis puts the "horse before the cart", with respect to fundamental thermodynamic drives and molecular structure-function correlation.

## Item 2: Evidences supporting murburn explanation for mOx-Phos

Herein, I shall not discuss any agenda that brings down the erstwhile hypotheses because such points have already been discussed in the manuscripts that debunked the prevailing explanations.

***Naturally self-assembled nuclear reactor:*** There is no historical evidence of a spontaneously formed automobile (with synchronized functions of engine, battery, dynamo, sensors/controllers, etc.) or hydroelectric power plant. (The logic of hitherto offered explanations for mOxPhos is analogous to these "working machine" models.) But earth does document solid evidence for a naturally assembled and spontaneously "initiated & sustained" nuclear reactor. The logic of the natural nuclear reactor minimally needed a nuclear fuel source (fissile uranium deposit), a neutron moderator (ground water) and an inlet for coolant (the same water!) outlet of pent-up entropy/energy (afforded by the steam being blown off!). Such a nuclear reactor did assemble/function spontaneously and completed a cycle every 3 hours (water enters deposit→ fission starts → water moderates neutrons → system temperature rises → water boils away → reaction slows/stops → system cools down). It worked cyclically for >$10^5$ years, until the fissile materials no longer had the critical mass necessary to sustain the reaction. The beauty of this finding was that the terms of feasibility and consequence of such an event was already predicted [Kuroda, 1956]. In 1972, Francis Perrin (a French physicist) made a chanced discovery of unusual distribution of various nuclides in Oklo mines, Gabon; thus validating the earlier prediction [Meshik et al, 2004]. Therefore, the operating principle based on a "nuclear reactor" model is a well-grounded and facile logic that is aligned with evolutionary pressures (because the constituting elements can be spontaneously formed and assembled) and Ockham's razor (because the elements are simple and constituted in straight-forward fashion).

***Why are DROS found in the mitochondrial systems?*** Experimental evidence shows that copious amounts of DROS are formed and found in real space and time, within the mitochondria in both normal and pathophysiological states. This fact, a reality, is wished away as "uncoupling" or "leak" or "artefacts" by researchers. Why should they be seen as mere pathological manifestations, if not for preconceived notions and "pseudo-aesthetics"? In some select instances, diffusible radicals' obligatory functional roles have been accepted. The much celebrated nitric oxide (NO, which serves as a molecular messenger) and enzymes like ribonucleotide reductase (which employs a metal centre that releases a confined diffusible radical for its routine functioning) have been around for a long time now! [Why shouldn't there be more paradigms like these in cellular systems? Why should superoxide generation (and peroxide, singlet oxygen and hydroxyl radical production thereafter) not be a spontaneous process within such systems?] The reality is that complexes I through IV of the respiratory pack do produce DROS (as exemplified by superoxide) [Grivennikova & Vinogradov, 2006; Drose, 2013; Bleier & Drose, 2013; Ksenzenko et al, 1992]. The "aesthetic/deterministic belief" system among some scientists that DROS would only be associated with chaotic and disruptive activities is nothing but superimposing their will over reality.

***Why is active cellular respiration and ATP synthesis synergic with the overproduction of DROS?*** It has been found in physiologically relevant states that production of superoxide (DROS) has been directly correlated to the in situ concentration of oxygen and the reduced substrates [Murphy, 1987]. Figure 1 from a wonderful review [Nicholls, 2004] clearly gives the direct message- maximum oxygen consumption (respiration rate) and maximum ATP formation (oxidative phosphorylation) is directly correlated to maximal ROS production within milieu. From the same figure, a clear correlation can be made that membrane potential only increases with the formation of DROS!

***Non-integral stoichiometry:*** Protons are stable and accountable species; their operations and stoichiometries in chemical reactions should given whole number and constant relations with the other participants within the system. While the researchers have been desperately trying to arrive at a consensus regarding this requisite, their experimental data seemed to speak the other way, whether it is P:O or H:P ratios [Rottenberg & Caplan, 1967; Lemasters, 1984; Hinkle 2005; Brand & Lehninger, 1977; Strotman & Lohse, 1988; Turina et al, 2003; Steigmiller et al, 2008; Petersen et al, 2012]. Such a scenario can be explained by invoking the obligatory involvement of ROS and their dynamics. Slight variations in experimental conditions can alter the ROS dynamics, which could give different non-integral values across various labs.

***Structures of proteins/complexes***: The matrix-ward projection of the five complexes are roughly ordered as follows- Complex IV (~3-4 nm) < Complex II (~8 nm) < Complex III (~8 nm) < Complex I (~14 nm) < Complex V (~15 nm) [Sazanov, 2015; Sun et al, 2005; Iwata et al, 1998; Tsukihara, 1996; Kuhlbrandt, 2015]. Only Complexes III and IV have any significant projections into the inter-membrane space. While the outward inter-membrane projections of Complexes III & IV can be explained by the prevailing hypothesis (so as to serve for binding the oxidized and reduced forms of Cyt. c respectively), the matrix-ward projections of Complexes I, II & III are quite a puzzle. Why did evolution not tuck away the redox centres of the flavoenzymes (Complexes I & II) into the transmembrane domain and choose not to send out finger projections into the matrix? Further, Complex III has no known substrate to bind at the matrix side, but yet, it has a big bulbous protrusion into the matrix. If the purpose is to minimize ROS, there was no need for having such a long finger projection into the matrix at all! If the erstwhile understanding held good, then the flavins should have been borne on a short-stub, to take the electrons efficiently and relay it to two chains of redox centres tucked away in the inner membrane (in an increasing order of potentials) so that electrons could be relayed along and protons be pumped out in the process! Or, it could just hand it over to CoQ directly and let the reduced $CoQH_2$ give out the protons at the IMS side through another simple functionality!! Why go over this long finger like structure in the matrix side (without any accompanying proton pumps!), only to invite trouble?!

***Distribution ratios/densities of the Complexes:*** The in situ / in vivo distribution of reaction components forms one of the major supports for the murburn concept. The reader can see from earlier published data [Gupte et al, 1984; Schwerzmann et al, 1986] that the mean distance between the two nearest proteins/complexes ranges between 7 – 20 nm and proteins/complexes are distributed at a density of $10^{10}$ to $10^{11}$ molecules per $cm^2$ (occupying totally a maximum of <50% surface area of the membrane). Figure 1 (of the main manuscript) shows a rough schematic for the relative distribution of the proteins/complexes involved. The ROS generators and moderators are at low concentrations (as expected!) and ROS stabilizer (Complex IV) is at high concentration. CoQ and Cyt. *c*, the soluble redox scavengers/buffers are at higher concentrations. Given the Complex IV : Cyt. *c* ratio at 1 : 1 to 1 : 1.5, the murburn concept seems more probable than any ETC-based system.

***Enzyme-catalyzed "non-active site" insertion of relatively non-oxidizable groups into final substrates:*** In chloroperoxidase catalyzed peroxidative chlorinations, chlorine atom insertion occurs through a comparable mechanism (at high rates!). It is known that chloride ions are very difficult to oxidize in mild aqueous milieu but when they are present at tens of millimolar levels, the chloroperoxidase enzyme mediates a "non-catalytic site" reaction. It releases a diffusible reactive species, which recruits the chloride ion, to non-chaotically and site-specifically chlorinate the final substrate via an electrophilic substitution [Manoj, 2006]. My recent work with cytochrome P450 shows that the hydroxylation of a xenobiotic substrate (an overall two-electron process) catalyzed by the radical intermediate, a one-electron species) is along the same logic [Manoj 2016d]. Again, in xenobiotic metabolism, non-specific addition of sulfate (and rarely, phosphate too!) is reported in liver microsomes [Mitchell, 2015]. These instances support my hypothesis that coupling could also occur through a non-active site, radical-mediated phenomenon. (If a specific sulfate or phosphate insertion enzyme was needed for all these diverse substrates, then we would need to start looking for another set of genome within all animals!) But it should also be noted that murburn concept is not against the involvement of any specific ADP binding sites (adjacent to the site of radical release), that could aid in this process.

***Re-interpreting the affects/effects of chemical uncouplers and uncoupling protein (UCP) of brown adipose tissue (BAT):*** If there exists a catalytic mechanism within the membrane to quickly protonate the superoxide formed, then we would have a loss of redox equivalents by futile cycles (that do not lead to an efficient ATP synthesis), and the process would lead to heat generation because the radicals would quickly amongst themselves. My group's works have shown that dihalophenolics (structurally similar to the "celebrated" dinitrophenolics that are known to uncouple ATP synthesis from NADH oxidation in mOxPhos) efficiently mess with the function of mXM by virtue of their interfacial DROS modulating abilities, not because of active site binding. Similarly, dinitrophenolics and other purported "proton shuttlers" [Terada, 1990] are in fact interfacial DROS modulators. The presence of SOD and catalase within the matrix cannot significantly mess with the DROS dynamics in the murzone near the internal mitochondrial membrane. But the N-term broken trans-membrane segment of CPR (that harbors positively charged amino acid residues and HRP, a DROS utilizing enzyme which has a transmembrane segment deleteriously affect DROS dynamics and CYP-mediated catalysis [Manoj et al, 2010b; Gideon et al, 2012; Parashar et al, 2014a,b; Manoj et al, 2016d]. Similarly, the UCP-1 of BAT has at least 9 positive extra positive charges, with several lysine and arginine residues lining the TMS and forming the matrix-side loop components [Klingenberg & Huang, 1999]. These could catalytically dismutate the superoxide and thus, mess with the DROS dynamics within the murzone.

***Toxicity of cyanide:*** Death by low amounts of cyanide poisoning does not result owing to the "binding" of cyanide to Complex IV because for the same, we would need at least a couple of orders higher cyanide concentration than the actual $LD_{50}$ that we currently observe. We have already presented evidence (in a reductionist system) that the ability of cyanide (taken at low concentrations) to inhibit heme enzymes is owing to diffusible radicals' involvement [Parashar et al, 2014b].

***Effects of cardiolipin and cationic lipids:*** The structure of cardiolipin shows a significantly higher negative charge density localized at the membrane periphery. This would facilitate superoxide retention within the matrix side of the murzone and thus, enable better ATP yields. On the other hand, cationic lipids (incorporated into liposomes for effective drug delivery) would have quite the opposite effect on superoxide, leading to dismutative functions. This would mess with normal physiology and prove detrimental to the cell.

***Apoptosis induced by cytochrome c:*** It is a well-known aspect of cellular physiology that presence of cytochrome c in cytosol coincides with the triggering of the apoptosis cascade. The simple logic that one could deduce is – mitochondria evolved to use DROS within the murzone. Once the integrity of outer membrane is breached, then the cellular components would be subjected to a load of redox activity, which could be intolerable to routine sustenance of cellular machinery for a multicellular organism. So, it is quite natural to assume that such a cell would program itself to a quick death, to minimize further harm to the rest of the cells.

***Coupling of ATP synthesis with proton input:*** In any reaction, one could envisage a catalytic effect and a participatory effect of certain key components. In a system in equilibrium, when starting with a particular ratio of reactants, the system comes to rest with x amounts of the product of interest (say, ATP). If the initial conditions and ratio of reactants are altered, the final concentration of the product of interest might get to a higher level than the earlier value. I would not call the latter any yield enhancement. Direct yield enhancement is when we start with the same concentrations of all the reactants and the overall reaction can be written without involving the catalyst. This is because the particular reaction component (catalyst) is not consumed in the process and thereby, it does not change the status of the equilibrium. When experimenters start with a lower pH outside and show that it results in a net synthesis of ATP, I don't think this deserves much attention! If in those conditions, the overall equilibrium is defined by the equation-

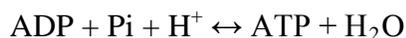

$$ADP + Pi + H^+ \leftrightarrow ATP + H_2O$$

Then, proton's addition to a confined (proton-limited!) system just displaces the equilibrium towards ATP synthesis.

***Why see the development of trans-membrane potential difference in respiring cells?*** An actively respiring mitochondrial membrane would have a significantly higher tran-smembrane potential because of the accumulation of hydroxide ion, superoxide radicals and (hydro)peroxide ions within the matrix side of murzone. It cannot be because of an excess of protons moving out, as Mitchell hypothesized. (Quite simply, protons are not a player in this regard. I have already shown the calculations justifying that statement in the manuscript that debunked the EPCR paradigm.) The charges on these ions can be neutralized by several modalities (as exemplified by calcium uptake or proton moving in).

### Item 3: Details of experiments predicted to validate/falsify murburn concept

***Toxicity studies:*** Besides binding to $Fe^{2+}$ or $Fe^{3+}$ species (of Complex IV), a toxic principle like CO or cyanide would also work in the milieu by the following reactions-

(i) Carbon monoxide poisoning

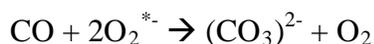
$$CO + 2O_2^{*-} \rightarrow (CO_3)^{2-} + O_2$$

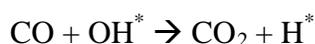
$$CO + OH^* \rightarrow CO_2 + H^*$$

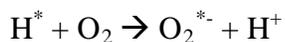
$$H^* + O_2 \rightarrow O_2^{*-} + H^+$$

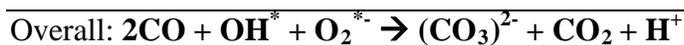
Overall: $\mathbf{2CO + OH^* + O_2^{*-} \rightarrow (CO_3)^{2-} + CO_2 + H^+}$

(ii) Cyanide poisoning

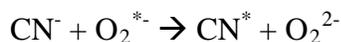
$$CN^- + O_2^{*-} \rightarrow CN^* + O_2^{2-}$$

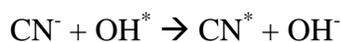
$$CN^- + OH^* \rightarrow CN^* + OH^-$$

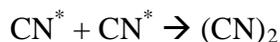
$$CN^* + CN^* \rightarrow (CN)_2$$

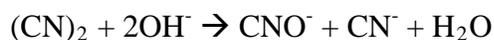
$$(CN)_2 + 2OH^- \rightarrow CNO^- + CN^- + H_2O$$

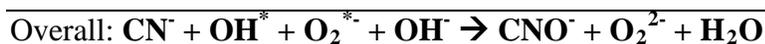
Overall: $\mathbf{CN^- + OH^* + O_2^{*-} + OH^- \rightarrow CNO^- + O_2^{2-} + H_2O}$

My predictions would be easily testable by dosing an experimental organism or system with radio-labelled CO & $CN^-$ (at significant but non-saturating concentrations!) and tracing the carbon in carbonate and cyanate formed. Kd of the protein should be determined at low concentrations of purified protein complexes and ligand. Then, in vitro assays can be carried out to determine the $IC_{50}$ at similar (or comparable) protein concentrations. Based on the logP/logD, a minimal formula can be used to project and correlate the maximum/minimum binding and inactivation *in vivo*. It would be seen that $K_d$ values cannot explain the $IC_{50}$ observed in both *in vitro* and *in vivo* scenarios.

***Uncouplers' maverick concentration-dependent effect*s**: Murburn concept would predict a concentration-dependent maverick modulation of electron transfer process, owing to chaotic reaction networks and partitioning effects in the phenomenology. (This could be why vitamin C is good to some and bad to others, with respect to curing or precipitating coryza like symptoms!) In reductionist systems, this affect/effect could be probed with mM to pM levels of uncouplers and I am quite confident that the researcher would get interesting results.

***Simulating the DROS-mediated coupling reaction and probing the effect of water activity and protons:*** Lipase catalyzed esterification is favored at low water concentrations (like water-in-oil emulsions and reverse micelles), and the opposite reaction is usually seen in oil-in-water emulsions. This is quite the reason why lipase-catalyzed esterification is favoured in reverse micelles. I don't see why it shouldn't be the same for phosphorylation of ADP (which is essentially an esterification reaction). The idea is that the free radicals generated in situ could catalyze the phosphate tagging to ADP. Therefore, the following reactions could be tried in simple aqueous milieu first (and luciferase enabled luminescence measured after a few minutes)-

1. ADP + Pi + NADH

2. ADP + Pi + Peroxide

3. ADP + Pi + Peroxide + NADH

4. ADP + Pi + Peroxide + Fe-Citrate

5. ADP + Pi + Peroxide + Iron-Porphyrin

6. ADP + Pi + NADH + Fe-Citrate

7. ADP + Pi + NADH + Iron-porphyrin

8. ADP + Pi + Peroxide + NADH + Fe-Citrate

9. ADP + Pi + Peroxide + NADH + Iron-Porphyrin

Buffer final ~ 90 mM (try pH 5.5, 7.5 & 9.5) Reactants-  ADP = 200 μM, Pi = 10 mM (final ionic strength is 100 mM total, with buffer), Peroxide = NADH = 200 μM; 10 to 100 nM catalysts. Iron porphyrin should be penta-ligated and could have midpoint potentials of ~ -250 mV and +150 mV. BSA could be used as a control.

I. A parallel set of reactions could incorporate DLPC vesicles of various dimensions (100 nm; 1 μm & 10 μm) and concentrations at ~ 1/10 μM, for the same combinations of reactants above. In these systems, small amounts of cytochrome P450 reductase could be added in conjunction with CYP2E1 + Cyt. $b_5$ or CYP3A4 alone. All reactions must have suitable controls.

II. Further,

    (i)      irradiating with a suitable UV light source could also be used to generate radicals from peroxide in situ or

    (ii)     superoxide cum peroxide could also be stabilized in DMSO + Crown ether and presented to ADP + Pi mixture.

III. Similar bunch of experiments (along with suitable controls) could be undertaken in AOT-isooctane reverse micelles at various water activity levels. (One must start with very low water amounts and establish the criteria for esterification first!) If water activity was important, ATP yield would lower with increasing micellar size.

IV. Murburn concept does not obligatorily seek proton pumping through $F_o$ component of Complex V for ATP synthesis. Therefore, it would be worth doing a simple reconstitution experiment (as per the protocol of Graber group, exemplified by Turina et al, 2003, at various pH ranges between 5.5 and 9.5; but in an even more reductionist setup) in which the following components, in the nine reaction setups sought above (under various circumstances, as detailed in the three points above)-

1. $F_oF_1$ ATPase

2. $F_o$ portion only

3. $F_1$ portion only

4. None

5. BSA

## Item 4: A comparative analysis of EPCR hypothesis and murburn concept

| No. | Attribute / Item | EPCR hypothesis | Murburn concept |
|---|---|---|---|
| 1 | | Nutshell of reaction paradigm | Multi-molecular ordered assemblies dictate a **vitally deterministic** course of events. Electron and proton separation over long distance and time across a phospholipid-embedded electronic circuitry leads to a proton/chemical gradient; this energy is tapped by a molecular motor enzyme to synthesize ATP. | At low proton and water activity, radicals generated in a lipid interface afford phosphorylation. Random collisions between molecules or ions or radicals afford high kinetic viability; equilibrium forces and **stochastic factors** (including spin barriers) dictate reaction fates. |
| 2 | *Chemico-physical aspects* | Outlay of reaction scheme | In the matrix side of the mitochondrial inner membrane, electrons are stripped from NADH and succinate, to be run down a "redox potential slope" (by a thermodynamic push); the energy released through some of these inter-molecular transfers is used to pump protons across the inner membrane (through a complex route within the proteins); then the protons' spontaneous flux back to the matrix is used to drive ATP formation by a molecular motor, which functions by conformation changes. | Flavo- and metallo-enzymes (in the presence of reducing substrates, oxygen and mobile redox active agents) set up a moderated relay of one- and two- electron reaction equilibriums within the mitochondrial inner membrane-matrix interface; formation of water and phosphorylated products serve as (small and bigger) two-electron sinks, pulling the overall reaction forward by a thermodynamic tug exerted. |
| 3 | | Coupling link between redox reaction and phosphorylation | No chemical connection, but a highly indirect electro-mechanical connection. | The formation of oxygen centered reactive intermediates directly aids phosphorylation. |
| 4 | | Electron transfer modality | Sequence of connected/charted events involving redox centres of proteins and small molecules in close proximity; movement/transfer of electrons primarily as pairs, via collisions or connectivity or supra complexes. | Randomized one- and two- electron reactions moderated at the membrane-matrix interface, with provisions for recycling of lost electrons. |
| 5 | | Thermodynamics & kinetics | Push of electrons from lower to higher redox potential centres; followed or synchronized with torque generation by a mechanized molecular motor (tapping into the generated transmembrane proton cum electro- | Shifting of one-electron equilibrium of oxygen-superoxide to the right by two-electron reactions occurring in the matrix side of the inner membrane. pH and ionic strength could alter the |
| | | | chemical gradient. | equilibriums. "Spin" could regulate crucial steps. |
| 6 | | Summation of overall redox reaction | $NADH + (C_4H_4O_4)^{2-} + H^+ + O_2 \rightarrow NAD^+ + (C_4H_2O_4)^{2-} + 2H_2O$ | $NADH + (C_4H_4O_4)^{2-} + O_2 \rightarrow NAD^+ + (C_4H_2O_4)^{2-} + H_2O + OH^-$ |
| 7 | *Biological aspects* | Evolutionary perspective | The whole machinery and process is aimed at reducing ROS formation. | The whole machinery is evolved for an effective utilization of DROS. |
| 8 | | Physiological outlook | Highly sophisticated machineries and processes set as pre-requisites for energy transduction mechanism. | Random and chaotic radical reactions serve to initiate & synthesize energy currency. |
| 9 | *Overall rationale* | Reducibility & Assumptions | Challenges Ockham's razor, A vectorial & complicated model that requires a sequentially ordered process. Proteins (and their complexes) need to have functional/mechanistic ability to sense and tap electrical potential and concentration gradients. Charge and mass transport across two macroscopic phases involved. | Aligns with Ockham's razor, minimalist model that has independent randomized events confined to the phospholipid interface. Molecules have predictable properties but the reaction scheme with such agents could be associated with significant levels of uncertainty. Reaction logic necessitates confinement of intermediates. |
| 10 | | Analogy | For simplicity: A water mill or hydroelectric power plant. For complexity: engine, dynamo & battery of an automobile. | For simplicity- an ordinary hearth. For complexity- a nuclear fission reactor. |
| 11 | | Oxygen | Involved in the end of ETC, as the terminal electron acceptor, bound at Complex IV alone | Reacts in the initial phase; involved in a multitude of reactions and equilibriums; omnipresent conduit |
| 12 | *Reactants* | NADH | 2e donor to Complex I | One or two electron reaction at Complex I and could have secondary roles; proton deficiency important |
| 13 | | ADP | Bound status to Complex V; required for ETC and phosphorylation | Not ascertained |
| 14 | | Pi | Bound status to Complex V; required for ETC and phosphorylation | Not ascertained |
| 15 | | Proton* | Used up on the matrix side; "pumped outward" and spontaneously moves inward | Involved in several reactions and equilibriums in the redox processes within the matrix |
| 16 | *Known catalysts (proteins &* | Complex I | *NADH dehydrogenase and Ubiquinone reductase & proton pump* Catalyzes two electron transfers from NADH to CoQ | *NADH oxidase* (a murzyme) Owing to its structural feature, proton deficiency within NADH and omnipresent oxygen molecule, |

| | | | |
|---|---|---|---|
| *complexes)* | | and takes one proton from matrix to supplement the proton shortage for $CoQH_2$ formation; pumps four protons out from the matrix; forms the "chassis" for respirasome. | provides one electron into the oxygen-superoxide equilibrium. CoQ can get one or two electrons into the lipid membrane from one or two NADH molecules' oxidation. |
| 17 | | Complex II | *Succinate dehydrogenase AND Ubiquinone reductase (& CoQ recycle proton supplier)* Catalyzes two hydride ion transfers to CoQ, forming $CoQH_2$ | *Succinate dehydrogenase, $FADH_2$ oxidase AND Ubiquinone reductase* Catalyzes the generation of superoxide, could also generate $CoQH_2$. |
| 18 | | Complex III | *Ubiquinone-cytochrome c oxidoreductase & proton pump* Intramembrane electron conduit, transferring two-electron equivalents from one molecule of $CoQH_2$ to two molecules of Cyt. c (accompanied by a very complex Q cycle!); pumps four protons out of the matrix. | *A murzyme with transmembrane bifunctionality* Projects that this component recycles the two-electrons lost into the membrane (to form $CoQH_2$ or CoQH) by transferring one electron equivalent to oxygen in the matrix side and one electron to Cyt. *c* in the inter-membrane space. Function needs to be re-defined/re-investigated. |
| 19 | | Complex IV | *Cytochrome c oxidase* Transfers 4 protons and four electrons to a long-term bound oxygen, to release two molecules of water; pumps four protons out of the matrix | *A complex murzyme* Recycles the electron lost to cytochrome *c* by transferring it to oxygen; and/or transfers it to the bound superoxide thereby reducing it to form peroxide; forms water because of superoxide + peroxide reaction in its vicinity (this gives a two-electron sink and a tug. |
| 20 | | Complex V | Driven by protons and works well in both directions, ATPase and ATP synthase modes. | Could serve to bring in protons and in driving equilibrium centered reactions. Primarily an ATPase, in the overall scheme. (Re-investiage!) |
| 21 | | Cytochrome *c* | One electron relay agent from Complex III to Complex IV. | Electron scavenger in the IMS; recycles by giving electrons off to Complex IV. |
| 22 | | CoQ | Diffusible 2e relay agent transferring two electron equivalents from Complexes I & II to Complex III | 1e & 2e redox agent across wide redox window |
| 23 | *Other key role players and stage makers* | Ions | Lead to electrical potential build up | Lead to charge relay facilitation and equilibrium effects |
| 24 | | ROS | Deleterious product formed in oxidative stress | Obligatory intermediacy of superoxide/hydroxyl radicals and agents like singlet oxygen and peroxide |
| 25 | | Cardiolipin | Structural component for respirasomes and renders ETC + proton pumping efficient, as it traps protons | Enables better conservation of superoxide equivalents within the murzone by high negative charge density |
| 26 | | Inner membrane | Selectively permeable, serves to maintain a (proton/chemical gradient) potential difference across itself | Serves to house and anchor murzymes that mediate radical reactions at the phospholipid interface |
| 27 | | Outer membrane | Present to concentrate protons | Present to prevent the escape of radicals |
| 28 | | Matrix | Reactions occurring within matrix are inconsequential to ETC or phosphorylation | Reactions within matrix govern the overall rates and outcomes |
| 29 | *Products* | ATP | Formed at Complex V by rotational catalysis by the mechanical push afforded by the inward movement of protons | Formed in matrix and more efficiently in the presence of Complex V, associated with protons' inward movement (other complexes' roles?) |
| 30 | | $NAD^+$ | Formed spontaneously at Complex I | Formed in the matrix/Complex I |
| 31 | | Fumarate | Formed spontaneously at Complex II | Formed spontaneously at Complex II |
| 32 | | Water | Formed at Complex IV and Complex V | Formed primarily in matrix |
| 33 | | Hydroxide ion | Not involved in the overall equation! | Generated in the matrix side of the inner membrane; serves to enhance longevity of hydroxyl radical |
| 34 | *Additives* | Inhibitors like rotenone, carboxin, etc. | Disrupt ETC by binding to Complex I & II | Take away activity of the component by altering the redox status of centre by covalent/affinity binding; preventing a 2e sink in vicinity |
| 35 | | Low amounts of inhibitors like cyanide, CO, etc. | Disrupt ETC by binding to Complex IV | Prevent the build-up of physiologically necessary radicals |
| 36 | | Uncouplers & UCP | Dissipate trans-membrane proton gradient | Disrupt ROS dynamics by facilitating superoxide dismutation at phospholipid interface The positively charged TMS of UCP disrupt ROS dynamics |